\documentclass[11pt,twoside]{article}
\usepackage{asp2010}

\resetcounters

\begin{document}

\title{The Past, Present and Future of Astronomical Data Formats}
\author{Jessica Mink$^1$, Robert G. Mann$^2$, Robert Hanisch$^3$, Arnold
Rots$^1$, Rob Seaman$^4$, Tim Jenness$^5$, Brian Thomas$^4$, William
O'Mullane$^6$
\affil{$^1$Smithsonian Astrophysical Observatory, 60 Garden St., Cambridge, MA~02138, USA}
\affil{$^2$Institute for Astronomy, University of Edinburgh, Blackford Hill, Edinburgh, EH9~3HJ, UK}
\affil{$^3$Office of Data and Informatics, Material Measurement Laboratory, National Institute of Standards and Technology, USA}
\affil{$^4$National Optical Astronomy Observatory, 950 N.~Cherry Ave,
Tucson, AZ~85719, USA}
\affil{$^5$Center for Radiophysics and Space Research, Cornell University, Ithaca, NY~14853, USA}
\affil{$^6$Science Operations Department, European Space Astronomy
Centre, European Space Agency, Madrid, Spain}
}

\begin{abstract}
The future of astronomy is inextricably entwined with the care and
feeding of astronomical data products.  Community standards such as FITS
and NDF have been instrumental in the success of numerous astronomy
projects.  Their very success challenges us to entertain pragmatic
strategies to adapt and evolve the standards to meet the aggressive
data-handling requirements of facilities now being designed and built.
We discuss characteristics that have made standards successful in
the past, as well as desirable features for the future, and an open
discussion follows.
\end{abstract}

\section*{Introduction and Data Format Basics (Jessica Mink)}
We are expanding the annual ADASS ``FITS BoF,'' which in recent years has
gone beyond that single data format, to include contributions to a special
issue of Astronomy and Computing on \emph{The future of astronomical data
formats}, which will provide a forum for peer-reviewed contributions to
the discussion of future data formats.

There are four major uses of data formats in astronomy, and it is
important to note that a single format does not have to fulfill all of
these purposes.

\emph{Recording}:   Instrument-specific, Metadata recorded.

\emph{Processing}: Software-specific, Metadata created.

\emph{Transferring}:  Well-documented, Metadata included.

\emph{Archiving}: Persistent and well-documented, Metadata included.

\section*{A Call to Action (Bob Hanisch)}
FITS is now about 35 years old, an eternity in the IT world, and we are at
risk of replicating the world of data format chaos that existed in the
late 1970s.  We should not be criticizing FITS for things that were not
possible when it was designed.  The time for complaining is over; who will
fill the roles of Harten, Wells, and Greisen \citep{1981A&AS...44..363W}?

A possible way forward, suggested by K. Shortridge, might be to use VO-agreed
data models \citep[e.g.][]{2012arXiv1204.3055M} as the high-level abstraction,
with HDF5 as the Processing and Transfer layer.

Should we retain FITS as the (an) Archive layer?  Perhaps we don't have
to solve all problems at once.  We should leave FITS otherwise alone so
as not to distract from a more general solution.

A timescale for action is that the IAU structure is being wiped clean
in a year. We need to get started now if we want to have anything
bear the imprimatur of the existing Commission 5.

\section*{FITS History (Jessica Mink)}

The FITS data format was developed to fulfill the basic needs of
human and machine readability, self-documentation, a ``universally''
readable format, and extensibility.  The presentation of the standard
FITS -- a Flexible Image Transport System \citep{1981A&AS...44..363W}
and later versions \citep[Definition of the Flexible Image Transport System
(FITS), version 3.0; ][]{2010A&A...524A..42P}
in refereed papers helped make its use widespread.

Over time FITS use has spread across categories, from Transferring to
Processing to Recording to Archiving, though maybe not in this order,
helped along by the motto, ``Once FITS, always FITS.''

Variations on the FITS format at NOAO and STScI were used as Processing
formats with machine byte order and separate data and metadata for ease
of processing. These turned into Transfer and to some extent Archive
formats, though they are convertible into FITS.

Gradually other standards have crept into FITS: world coordinate systems
for space, spectroscopy, and time, binary and ASCII tables, and multiple
extensions in a single file.  A registry exists to document site-specific
keywords.

\section*{FITS Time Paper (Arnold Rots)}

From the start FITS has provided a well-defined grammar and syntax for
writing astronomical data, but the standard did not include a semantic
component. In many respects that was a good thing, since it would have
been impossible to develop an all-encompassing semantic vocabulary.
However, it is not sufficient for readers to be able to read the data;
they also need to be able to understand them.  In practice, this has
been solved by sub-communities developing their own conventions. The
solution works well, as long as there is not too much overlap between
those sub-communities. But it became apparent very soon that the need
to record and transmit coordinate information is common to all, and this
resulted in the development of common World Coordinate System (WCS)
standards. The first two papers in this series dealt with general
principles, spatial coordinates, and projections. The third paper
provided the standard for spectral, redshift, and Doppler velocity
coordinates. This past summer the IAU FITS Working Group approved the
fourth WCS standard, on the time coordinate, and the paper
\citep{2014arXiv1409.7583R} was accepted  by Astronomy \& Astrophysics
a week before the ADASS conference.  One may consider this paper the
concluding part of the WCS standards and, possibly, the final piece
of the FITS standard.

\section*{FITS Long-term Evolution (Rob Seaman)}

Let us make a rough estimate of the world-wide FITS data
holdings. There have been 15 million FITS files archived at NOAO over
20 years; due to large multi-chip cameras this is 50 million FITS
IMAGE HDUs . There are many hundreds of ground-based O/IR telescopes
compared to about ten at NOAO and we can conservatively say that there
are at least twenty NOAO equivalents in large and small, public and
private observatories around the world. Twenty times fifty million is
one billion FITS images. This does not include the tallies for radio
and space observatories, etc., so even if NOAO has been more
productive than most the estimate should be in the ballpark.

All agree that FITS has been very successful. FITS offers permanence
(for example the Vatican manuscript project). Archival FITS holdings
are extensive and growing and converting formats would be hugely
expensive; see also ``Data engineering for archive evolution'' from this
conference \citep{P3-2_adassxxiv}. For all these reasons and
more, support for FITS must continue under any future file format
scenario. We can make near-term enhancements if we choose; some which
the FITS Technical Group have been discussing include longer keyword
names, longer string-typed keyword values, and expanded character set
for headers.

But more importantly we must identify a strategy for FITS to continue
to represent the full richness of evolving data formats and data
models in astronomy. The original FITS achieved the enviable goal of
serving as astronomy's lingua franca. FITS can / should / must
continue to serve its original roles of data transfer and data
archiving. The capital-T in the name stands for Transport: transport
in space and transport in time. There is enough flexibility in the
binary table paradigm of FITS to be able to support any semantically
rich data structure that can be represented as a table (for one simple
example, to capture a FITS image header with enhanced keyword
attributes). It may be that there are new data recording or data
processing use cases for which FITS images and binary tables are no
longer sufficient, but many other future astronomical niches will
require nothing other than a more refined use of current FITS
capabilities.

\section*{Astronomy and Computing special issue (Bob Mann)}

Astronomy and Computing \citep{2013A&C.....1....1A} are preparing a
special issue on \emph{The future of astronomical data formats}.
Summaries of two papers from the issues are being presented now.
Further papers are in preparation, and it is hoped that this BoF and
subsequent discussion this week will generate more.  The final
submission deadline is 1 March 2015.  All papers from the Special
Issue are being posted on the Astronomy and Computing
website\footnote{\url{http://www.journals.elsevier.com/astronomy-and-computing}}
in preprint form to enable inclusion in the debate.  There are three
so far:

Tim Jenness et al: Lessons learned from NDF \nocite{2015A&CNDF}

Slava Kitaeff et al: Use of JPEG2000 for astronomical imaging \nocite{2014arXiv1403.2801K}

Brian Thomas et al: Learning from FITS \nocite{2015A&CFITS}

\section*{Moving NDF to HDF5 (Tim Jenness)}
The extensible N-Dimensional Data Format \citep{2014ASPC..485..355E,2015A&CNDF}
is a data model developed in the late 1980s to solve the
problem of unbounded expressiveness supported by a hierarchical data
format. The model provides a framework for placing information such as
data, variance, quality, world coordinates and history into a file. NDF
is implemented as a Fortran library layered on top of the Starlink
Hierarchical Data System (HDS). HDS is a hierarchical file format
developed in the early 1980s and currently in version 4. It is written
in C but is no longer supported by anyone who understands the complex
file structure and implementation details. In order to broaden support
for NDF in the community HDS version 5 will be a reimplementation of the
HDS API but using HDF5 as the underlying file format. This will allow
NDF and its associated library to be used by others without taking on a
niche unsupported file format. Currently a prototype library has been
developed which can create and query files using the HDS API. The final
aim is to produce a library that can transparently read and write
version 5 files written in HDF5 but also read older format files, the JCMT Science Archive \citep{2015A&CJSA}
contains more than a million such files, and allow them to be migrated to the new format.

\section*{Using FITS to understand astronomical data format 
needs (Brian Thomas)}

ITS is a great ``test particle'' for analyzing astronomical data format
needs. There are many reasons which include the fact that it is the
Lingua Franca of astronomical data, it is well-documented and tested and
has many technical strengths along with good software support. 

Building on our previous work \citep{2014ASPC..485..351T,2015A&CFITS}
our process is to form a large group of individuals of varying backgrounds
to collect and analyze issues on the astrodataformat Google group and
associated Github organization\footnote{\url{https://github.com/astrodataformat}}.
We invite a wide variety of people to participate, but set and enforce
ground rules to lower chances of acrimony and focus our discussion.
Our goal is to reach a consensus view and share our results with the community.

In our recent work which focused on deficiencies within the FITS standard
we find that problems may be grouped into 2 categories.  First, there are
the well-known limitations which include  its metadata expression
(8 char keywords, 68 char values, hierarchical structures not 'native', no
built in associations), data model issues (inflexible WCS, associations),
and serialization (choice of endian, missing values).
In the other category there are the new needs which have surfaced over time.
New needs include things such as the need for greater exchange of data both in
terms of the amount of files and the number of bytes per file.
These new needs in turn create new requirements/demands on the format such as
an increased need for validation and machine understanding, virtualization/distributed
support and more and improved data models.

A single standard for sharing data is a \emph{huge} boon for the astronomical
community, \emph{but} FITS is showing its age.  If we want to continue having
this kind of shared standard, then FITS needs to evolve sufficiently or
a new standard needs to be found.

But how? Should we choose to evolve through existing standard/conventions,
apply radical surgery, create a new data format which translates
the useful data models of FITS \citep[e.g.][]{O4-4_adassxxiv,P3-1_adassxxiv}, 
or perhaps start completely from scratch? Because we want a shared a community 
standard, we feel strongly the only means to achieve this is by engaging the 
community.

We plan to continue our work and plan to expand our scope in the next stage
by gathering use cases and ``lessons learned'' which also show FITS strengths,
and gleaning the same from other data formats. From these we will extract
requirements and we will use a voting process to allow the group to determine
priorities and eliminate edge cases which are of little value. We hope our next
effort can help to inform the community and provide a starting point for
the development of a astronomical data format for the future.

\section*{Discussion}

\emph{Frank Valdes}: Modifying APIs. What APIs?
\\
\emph{Tim Jenness}: will rip out guts of HDS library and replace by HDF5. Would
provide wrapper to use NDF, with HDF5 as the underlying file format.
Others can use NDF without worrying about HDS.
\\
\emph{Frank Valdes}: two paths -- modifying FITS or going to something else, so
need to consider the former.
\\
\emph{Bob Hanisch}: some of the major drivers are data size and patches for FITS
are really just patches, and don't address the more fundamental issues
with FITS, so would be a distraction to patch FITS only.
\\
\emph{Frank Valdes}: data size raised as a concern, but I can work with 5GB MEFs.
What is it about size?
\\
\emph{Bob Hanisch}: HDF5 can stream sections of files across multiple spindles,
has high spec I/O. Concern with HDF5 is archival -- defined by API not
description of how the bits are arranged on disk.
We could make FITS more like HDF and still fall short of what is required.
\\
\emph{Jessica Mink}: FITS's simplicity has allowed it to take a market share for
all four types of data formats, but new formats need not. FITS is good
for archiving. For processing, recording and transfer  it would be nice
to have a standard format. Extending FITS in a standard way is very
supportable, and could then read old files very easily. FITS doesn't have
to be everything.
\\
\emph{Tom McGlynn}: QWERTY is not optimal, but it is a long-lived standard, so
there would be a high cost of changing. Is it the same for FITS?
It is hard to see the archival issue being solved by HDF5. Then we need
to think about conversion.  We need to take over all four types of format,
or transform between them a lot, and need to understand how to do that.
\\
\emph{Tim Jenness}: Starlink software converts to FITS from NDF all the time, so we
worked out how to flatten hierarchical data models to flat FITS, so we
can do that -- but it is painful, and currently requires convention.
It would be good to have a standardized way to represent hierarchy in
the FITS world, but not so big a problem as you might think
\\
\emph{Brian Thomas} refuted the archivability of FITS. He has FITS files stretching
back decades. FITS allows syntactic validation, but not semantic validation,
so meaning is not persisting in the same way over time. We need to have
documentation in the format, not in a paper or manual and versioning
for an archival format. We can read FITS files, but can't use them for
science if we can't understand them.
\\
\emph{Rob Seaman}: Can represent schema in a binary table, with a
version as an attribute – so can already implement semantic metadata
in FITS.
\\
\emph{Anne Raugh}, who is working on v4 of the Planetary Data System standards \citep{P3-3_adassxxiv},
noted that one needs to take step back and see what the problem is you now
need to solve, not evolve a solution to an old problem, but to new ones.
Having reached that stage with PDS, we decided to make a break.  It was
harder to evolve than to start again to address the current problems.
There are a number of pressures on FITS, in the Small Bodies Node of PDS,
because the format is useful. Fundamental changes to the FITS header
should be made cautiously. Then there is the problem of migrating legacy
FITS data to the new standard, but that's really just a metadata
problem - the data structures are very stable
\\
\emph{Ken Anderson}, who worked on LOFAR ICD specifications, agreed with the
previous speaker's statement. The misnomer is introduced by HDF itself.
HDF is not a format; it's a framework allowing people to define formats.
LOFAR has five or six formats under HDF5 \citep[see e.g.][]{2012ASPC..461..283A}. Specification  was by the
astronomical community, as with FITS. LOFAR defined an initial spec for
data products, some very complex and hierarchical, so a recognised
hierarchical data format was required.  HDF defines only an API; the
community needs to define the formats for the data. LOFAR ICDs are
available on LOFAR website. Because HDF is not a format by itself, it
is not an archival format.
\\
\emph{Tim Jenness}: NDF created to go on top of HDS because earlier anarchy in
use of HDS. NDF arose as minimalist data model required to do science. Two
options: use complicated VO data model or use NDF, which covers the basics
and leaves scope for extensions. Data models -- where do you go for them?
The LSST pipeline uses databases, not files on disk.  This whole debate is a
distraction from discussion of data models.
\\
\emph{Peter Teuben?} commented on the historic value of FITS. Astronomy has been lucky
to have a standard format, since we can build applications without bothering
with data ingestion.  Every experiment at LHC starts from scratch coding
the data ingest. Astronomy applications are more standardized. Separation of
metadata and data began many years ago and is very valuable. It is important
to keep that.  Semantics worry people. Expression of observatory keywords
does not use a controlled vocabulary, but we can now use VO standards for
metadata -- defined vocabularies (UCDS, etc). Metadata in VOTable with
embedded FITS for data load -- why not keep that sort of model for the
future? FITS became standard because coolest kid adopted it. Expect
that the next cool kid -- LSST, SKA -- will determine what we end up using.
\\
\emph{Tom McGlynn}: HST had to be pulled back to standard FITS -- it wasn't the
cool kid that led to the adoption of FITS.
\\
\emph{Bob Hanisch}: if had LSST, JSWT, ALMA agreeing to something, others would probably
follow -- not quite the history of FITS. NASA decided before HST was launched.
\\
\emph{?}: how did adoption in NASA lead to wider adoption?
\\
\emph{Jessica Mink}: AIPS was FITS-based -- availability of software led to adoption of FITS. Took over optical when IRAF started using it.
\\
\emph{Bob Hanisch} pushed back on Tim: IVOA hasn't created complex data models.
Strive to have minimum scope to function -- so all is under control!
\\
\emph{Francoise Genova}: UCDs vocabulary -- someone checked lots of FITS files
to check that the quantities in the FITS files were included in the UCD.
Only a few quantities were unclear in meaning.
\\
\emph{Rob Seaman}: FITS had a history before NASA: NRAO and KPNO created
it originally.
originally. Perhaps can solve metadata problems with IVOA data models,
but don't forget that there are data that go along with that. For
instance, could have HDF file with 8 different kinds of compression;
must be able to translate this to and from FITS. The logistics of FITS
tile compression have already started down slope to new formats
\citep{2007ASPC..376..483S}. Must remember that there are data as
well as metadata. Efficient data representation is key \citep{2010PASP..122.1065P}.
\\
\emph{Peter Teuben}: data model is the important thing, not how the bits are
stored on disk.  Companies have looked at objects, not files. Astronomy
could be screwed if IT industry moves away from files.
\\
\emph{Brian Thomas}: have to get away from storing information in a file -- need
to have data format extract information from database, etc. What's important
is the information not the file -- maybe applications are more important?
It is if it generates a subset of data when needed.
\\
\emph{Peter Teuben}: but we complain about HDF being an API not a format
\\
\emph{Brian Thomas}: We need to understand what we need -- gather use cases, extract
requirements, prioritise them and then maybe there will be four or two
data formats in future, but want to led by requirements from the community,
not seeking buy-in for a particular solution.

\acknowledgements
Thanks to Will O'Mullane for keeping us all on time so that everyone who
wanted to got to contribute and to Bob Mann for taking very good notes
on the discussion.

\bibliographystyle{asp2010}
\bibliography{ADASS2014B1}

\end{document}